# Calculation of point defects in rutile $TiO_2$ by the Screened Exchange Hybrid Functional

Hsin-Yi Lee,[1] Stewart J. Clark,[2] and John Robertson[1,*]

[1]*Engineering Department, Cambridge University, Cambridge, CB2 1PZ, United Kingdom*
[2]*Physics Department, Durham University, Durham, DH1 3LE, United Kingdom*

The formation energies of the oxygen vacancy and titanium interstitial in rutile $TiO_2$ were calculated by the screened exchange (sX) hybrid density functional method, which gives a band gap of 3.1 eV, close to the experimental value. The O vacancy gives rise to a gap state lying 0.7 eV below the conduction band edge, whose charge density is localised around the two of three Ti atoms next to the vacancy. The Ti interstitial generates four defect states in the gap, whose unpaired electrons lie on the interstitial and the adjacent Ti $3d$ orbitals. The formation energy for the neutral O vacancy is 1.9 eV for the O-poor chemical potential, and similar to that of the neutral Ti interstitial, and has a lower formation energy for Ti interstitial under O-rich conditions. This indicates that both the O vacancy and Ti interstitial are relevant for oxygen deficiency in rutile $TiO_2$ but the O vacancy will dominate under O-rich conditions. This resolves the questions about defect localisation and defect predominance in the literature.

## I. INTRODUCTION

Titanium dioxide ($TiO_2$) is a transition metal oxide with a closed shell electronic structure and a band gap of 3.05 eV. It is an important material for photocatalysis, solar cells and environmental cleanup, due to unusual photocatalyic behavior [1-4]. Oxygen defifiency defects such as surface oxygen vacancies are held to be critical for such applications. $TiO_2$ is also interesing in electronics. It has a high dielectric constant but its band offset is too low for it to be useful as a high dielectric constant gate dielectric [5]. On the other hand, it is of considerable interest for use in non-volatile resistive random access memories (RRAM) [6,7]. The conductive track in the on-state is believed to arise from a percolation path of oxygen vacancies or similar defect across the film. Thus, the behavior of excess electrons and associated defects is critical to the performance of $TiO_2$ as a catalyst or in electronics.

It is known from ultraviolet photoemission spectroscopy (UPS) and electron energy loss spectroscopy (EELS) that the oxygen deficiency defects give rise to gap states lying about 0.7 to 1.0 eV below the conduction band edge [8-13]. However, the origin and localisation of these states are still contentious. Electron spin resonance (ESR) showed that unpaired electrons associated with the oxygen vacancy are localised on two of the three Ti sites adjacent to the vacancy [14]. For many years, catalytic activity has been discussed in terms of surface oxygen vacancies, based on resonant photoemission or EELS. However recently, Wendt *et al.* [15] interpreted scanning tunneling microscopy (STM) maps in terms of a dominant role of Ti interstitials. Mass transport during $TiO_2$ growth also suggested that Ti interstitials were involved [16]. This debate has continued, with recently resonant photoemission spectra again suggesting that localised gap states were due to the oxygen vacancy [17-19].

Electronic structure calculations of these defects should be able to define the dominant defect and say how localised their defect states are. There have been many calculations of the defects of $TiO_2$ using the local density approximation (LDA) or generalised gradient approximation (GGA) [20-23]. However, LDA and GGA are known to under-estimate band gaps, and give defect states that are too delocalised, because of a lack of self-interaction cancellation [24,25]. This failure is very important for a material like $TiO_2$ where the defect states are on the border line between shallow and deep, and where the degree of localisation is the key question.

One method to correct these problems is the LDA+U method, in which an on-site repulsion U is added to the transition metal *d* states [26-31]. This can widen the band gap of $TiO_2$ slightly, but LDA+U should really be only used for open shell systems. On the other hand, LDA+U does give a better description of defect localisation, and in fact Morgan and Watson [27] found that U ~ 5 eV gives a reasonable description of defect localisation even if the band gap is still under-estimated compared to the experimental value of 3.05 eV [32].

Pacchioni [25] has discussed the need for the correct description of defects in the catalysis problem, and the problems with various electronic structure methods. The defect localisation in $TiO_2$ comes not only from the electronic state, but also the interaction of electronic states with lattice distortions – a polaronic effect, as noted generally by Lany and Zunger [33]. Hybrid density functionals are a better method to tackle such cases, because they combine a much improved description of the exchange energy and, as generalised Kohn-Sham functionals, they can be used for total energy minimisation and structure relaxation. Di Valentin *et al.* [34,35] used the earliest hybrid functional B3LYP to describe electron excess defects in $TiO_2$ and compare the results to LDA+U. This work provided a good understanding of the general defect properties, where the subtle energy differences between localised and delocalised defect states became apparent. Their results required some interpretation because their B3LYP over-estimated the $TiO_2$ band gap at 3.9 eV. In contrast, Muscat *et al.* [36] found a gap of 3.4 eV in their B3LYP results.

Janotti *et al.* [37] and Deak [38] have employed the widely used hybrid function HSE06 functional [39,40] to study the electronic properties of the oxygen vacancy and substitutional dopants. They found that HSE gave a greatly improved description than GGA, but they did not treat the Ti interstitial. They found the vacancy to have localised states in the gap but that the transition level was shallow.

A further improvement on the various methods might be posible using the GW method, but this would lose the simplicity of a single shot approach to structural relaxation for polaronic defects. We therefore here employ the screened exchange (sX) hybrid density functional [41-44] to treat the vacancy and interstitial defects. The sX method has been found to give the correct band gaps of a wide range of semiconductors such as the III-Vs, ZnO and $SnO_2$ and the insulators $HfO_2$ and $SiO_2$ [43], and the correct localisation of holes in the Al doped $SiO_2$ (smoky quartz) [45] and at the Zn vacancy in ZnO [46]. It also describes correctly the correlated systems such as the transition metal oxides $Ti_2O_3$, $Cr_2O_3$, $Fe_2O_3$, and the lanthanide oxides $Ln_2O_3$ [47] and NiO, FeO and MnO [48]. Its better description of the lanthanide oxides than HSE arises from its slightly better treatment of the highly localised $4f$ states, and this might be valuable in the present case where defects are on the shallow-deep border line.

## II. METHODS

The sX method is a hybrid density functional which replaces part of the short range part of exchange in LDA with a Thomas-Fermi screened non-local exchange as in Hartree-Fock. It can therefore be self-interaction free. The exchange term is constructed to satisfy both the low electron density and the free electron gas limits. The screened-exchange potential is expressed as

$$V_{sX}(r,r') = -\sum_i \frac{\psi_i(r) e^{-k_{TF}|r-r'|} \psi^*_j(r)}{|r-r'|}, \qquad (1)$$

where $i$ and $j$ label the electronic bands. The $k_{TF}$ is a Thomas-Fermi screening length. This can be set to the average valence electron density of the system, or given by a fixed value. Here, we set $k_{TF}$ to 2.15 Å$^{-1}$ by considering the experimental value of the band gap of rutile $TiO_2$. sX belongs to a class of generalised Kohn-Sham functionals so it can be used variationally for energy minimisation, using the calculation of forces [43].

The calculations were performed with the CASTEP plane wave pseudopotential code [49]. The cores were represented by norm-conserving pseudopotentials, while the valence states were expanded in a plane-wave basis set with 750 eV cutoff energy. For k-point, we used the Γ point in defect supercells. The geometry optimisation used the Broyden–Fletcher–Goldfarb–Shannon (BFGS) algorithm with convergence to 5 x 10$^{-5}$ eV per atom, 0.1 eV/Å for the Hellmann-Feynman force on each atom and a stress of 0.2 GPa.

The defect calculations use 2 x 2 x 3 supercells of 72 atoms plus the defect. A series of convergence tests had been applied to determine the computational parameters, including the cutoff energy, k-point and supercell size. Because of the polaronic nature of the defect, the defect supercell was relaxed in sX to obtain the correct lower symmetry geometry. A larger supercell of 3 x 3 x 4 units (216 atoms) was calculated to check the finite-size effects. We found the difference of O vacancy formation energy between 72 and 216 atoms is only 0.12 eV, so the sX calculations on a 2 x 2 x 3 supercell are adequate. We used spin-polarisation for the odd charge states to describe the unpaired electrons; for even charge cases, we have checked the results with and without spin-polarisation, and the data show negligible difference. For comparison, we also carried out the GGA-PBE calculations for the band gap which implied the ultrasoft pseudopotential with 380 eV cutoff energy in 5 x 5 x 8 k-point mesh, to treat as a reference to compare with the sX results.

For the calculation of defect formation energy in the different charge states, the overall supercell size was kept fixed at the relaxed neutral bulk value. The defect formation energy ($H_q$) of charge $q$ as a function of the Fermi energy ($E_F$) and the chemical potential ($\Delta \mu$) of element $\alpha$ is given by [50]

$$H_q(E_F, \mu) = [E_q - E_H] + q(E_V + \Delta E_F) + \sum_\alpha n_\alpha (\mu_\alpha^0 + \Delta \mu_\alpha) \qquad (2)$$

where $E_q$ and $E_H$ are the total energy of a defect cell and a perfect cell respectively calculated of charge $q$, $\Delta E_F$ is the Fermi energy with respect to the valence band edge, $n_\alpha$ is the number of atoms of element $\alpha$, and $\mu_\alpha^0$ is reference chemical potential, following the method described by Lany and Zunger [50].

There are three polymorphs of $TiO_2$, rutile, anatase and brookite. Here, we study the most common phase, rutile. $TiO_2$ rutile has a 6-atom tetragonal unit cell with the $P4_2/mnm$ space group. This lattice consists of 3-fold coordinated oxygen atoms and octahedral Ti atoms. Figure 1 and Figure 2 show a unit cell of rutile $TiO_2$ and the crystal structures of O vacancy and Ti interstitial where the defect sites are indicated by the arrows.

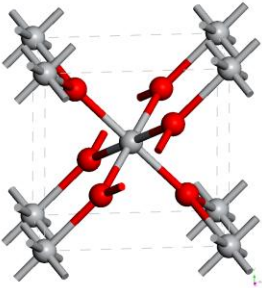

FIG. 1. (Color online) A unit cell of rutile $TiO_2$ where the red spheres represent O atoms and the grey spheres represent Ti atoms.

(a) O vacancy  (b) Ti interstitial

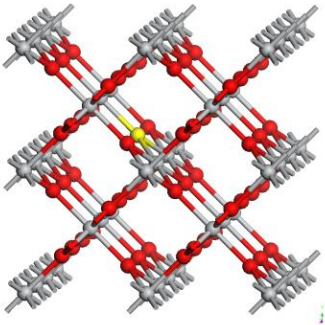 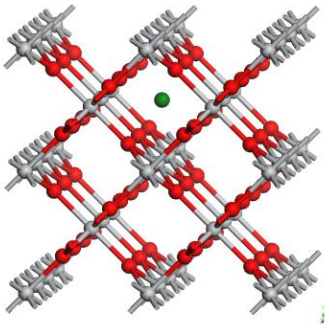

FIG. 2. (Color online) The crystal structures for rutile $TiO_2$ of (a) O vacancy, and (b) Ti interstitial. The defect site of O vacancy and Ti interstitial are marked by yellow and green spheres respectively.

## III. RESULTS

### A. Bulk

The band structures of rutile TiO$_2$ derived from the GGA-PBE and the sX are shown in Figure 3(a) and 3(b), respectively. The optimised structural parameters, the band gaps, valence band width, and the formation enthalphy ($\Delta H_f$) obtained with the GGA-PBE and sX methods are given in Table I together with the corresponding experimental values. [32,51,52]. We found that the band gap increases from 1.86 eV in GGA-PBE to 3.1 eV in sX, very close to the experimental value of 3.05 eV. The valence band width increase from 5.65 eV in PBE to 6.27 eV in sX, which is closer to the experimental value of 6.0 eV [52]. The heat of formation changes from −9.33 eV in PBE to −9.73 eV in sX, the latter is very close to the experimental valuie of −9.74 eV. The a lattice constant changes from 4.65 Å in PBE to 4.56 Å in sX, which is now slightly below the experimental value of 4.59 Å.

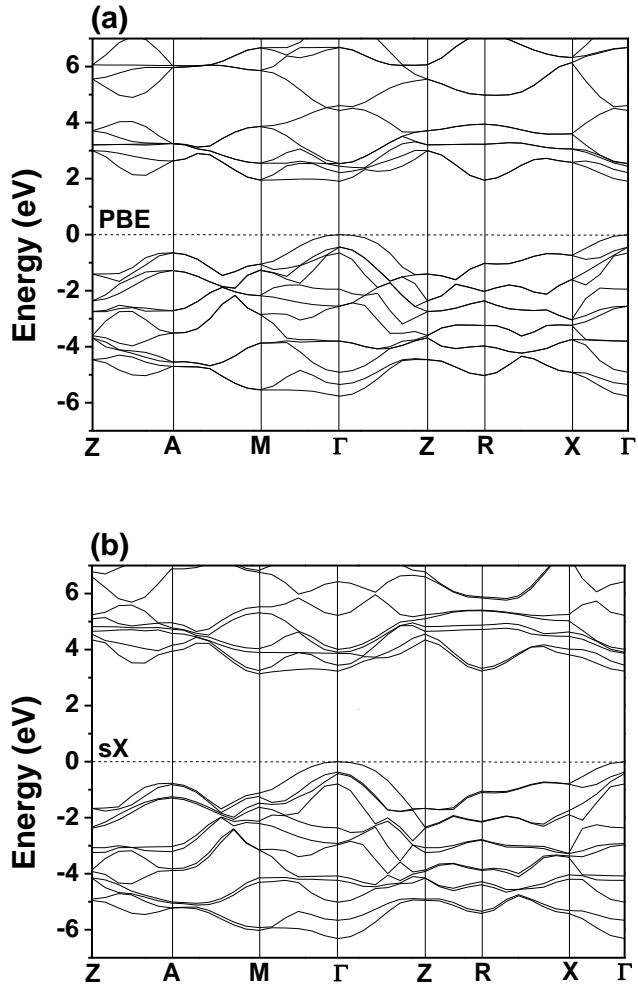

FIG. 3. Band structures of rutile TiO$_2$ calculated by (a) the GGA-PBE and (b) the sX methods.

|  | PBE | sX | Experiment |
|---|---|---|---|
| Lattice constant, $a$ (Å) | 4.65 | 4.56 | 4.59 |
| $c/a$ | 0.639 | 0.648 | 0.644 |
| $u$ | 0.305 | 0.305 | |
| Heat of formation $\Delta H_f$ (eV) | -9.33 | -9.73 | -9.74 |
| Band gap (eV) | 1.86 | 3.1 | 3.05 |
| VB width (eV) | 5.65 | 6.27 | 6 |

TABLE I. Structural parameters, heat of formation, minimum band gaps and valence band widths of rutile TiO$_2$ from GGA-PBE, the sX method and experiment [32,51,52].

### B. Oxygen vacancy

We first consider the oxygen vacancy (V$_O$). Fig. 4 shows the plane view of ion-ion spacings around the ideal vacancy. Removing one oxygen atom from TiO$_2$ results in two unpaired electrons and three Ti dangling bonds. The ideal spacings between adjacent Ti atoms are two at 3.55 Å and one along the $c$ axis at 2.99 Å. We denote V$_O^0$, V$_O^+$ and V$_O^{2+}$ for the neutral, singly charged and doubly charged states respectively for the oxygen vacancy. For V$_O^0$, after relaxation the adjacent Ti atoms move outwards, so the Ti-Ti distance increases to 3.7 Å, and 3.04 Å along the $c$ axis. For the singly positive vacancy V$_O^+$, the positive charge causes a greater repulsion for a Ti-Ti distance to 3.85 Å and 3.12 Å along O$_z$. For V$_O^{2+}$ the Ti-Ti distance increases to 3.94 Å and 3.22 Å along O$_z$. In other words, the lattice relaxation is 4.2 % for V$_O^0$, 8.5 % for V$_O^+$ and 11 % for V$_O^{2+}$. A similar outward relaxation is found for the O vacancies in other ionic oxides such as HfO$_2$ and ZnO [44,46].

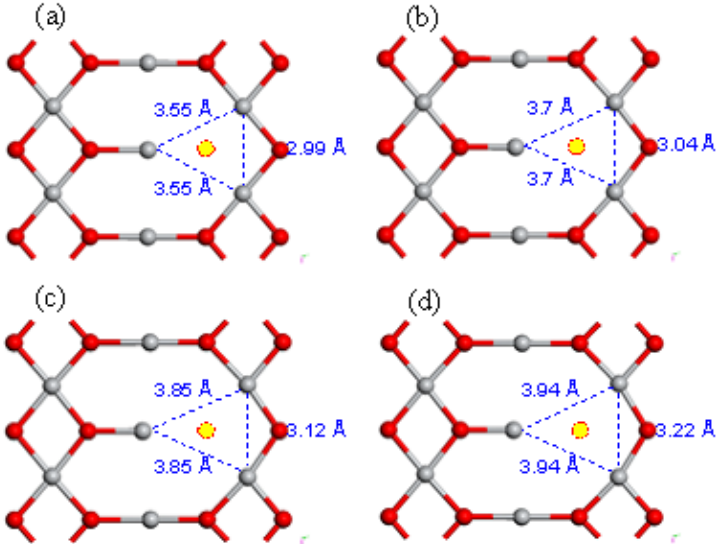

FIG. 4. (color online) Plane view of ion-ion spacings around the O vacancy site (yellow circle) after the geometery optimization carried out by the sX functional, (a) Before lattice relaxation; (b) V$_O^0$ relaxed; (c) V$_O^+$ relaxed and (d) V$_O^{2+}$ relaxed.

The defect formation energies can be calculated as a function of O chemical potential. The chemical potentials satisfy $\mu_{Ti} + 2\mu_O = H_f(TiO_2) = -9.73$ eV (experimental value). The O rich limit is $\mu_O = 0$ eV and $\mu_{Ti} = -9.73$ eV, correspoding to the chemical potential of the $O_2$ molecule. The Ti-rich limit corresponds to the equilibrium of $TiO_2$ and $Ti_2O_3$ (not $TiO_2$ /metallic Ti as in many cases) or $\mu_O = -4.07$ eV, $\mu_{Ti} = -1.59$ eV. The metallic Ti / $TiO_2$ equilibrium would correspond to $\mu_O = -4.86$ eV and $\mu_{Ti} = 0$ eV.

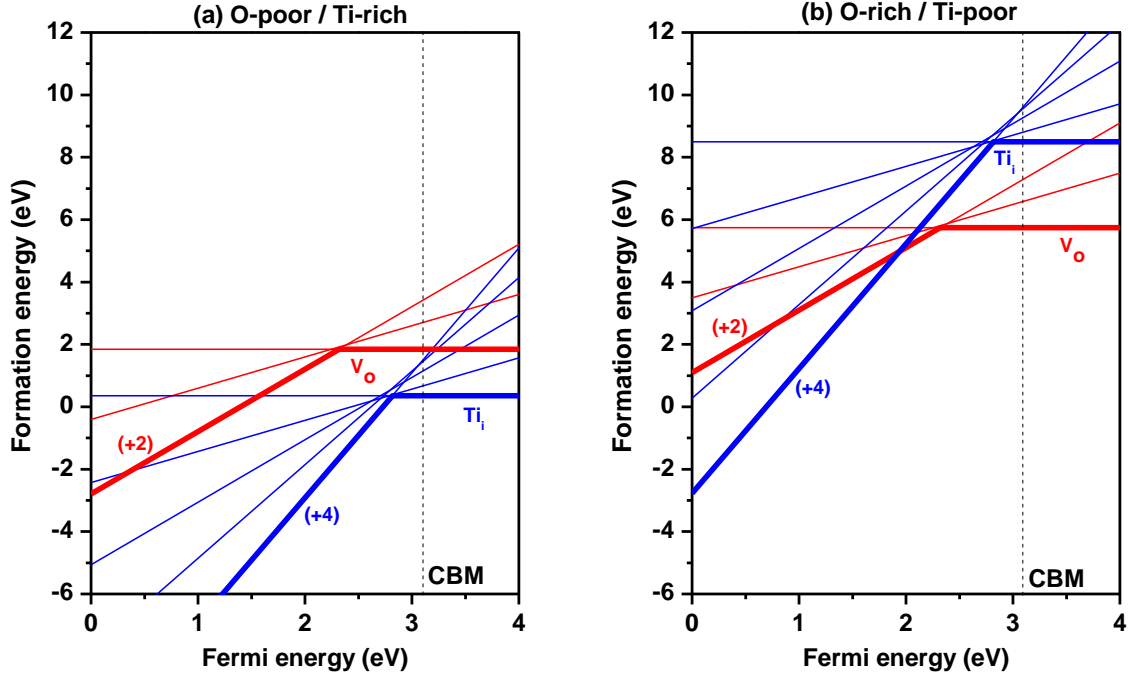

FIG. 5. The defect formation energy against Fermi level of the O vacancy (red lines) and Ti interstitial (blue lines) under (a) O-poor and (b) O-rich conditions, calculated by the sX functional for rutile $TiO_2$. The vertical dash lines denote the conduction band minimum (CBM). The slope of the formation energy lines gives the charge state of the defect.

Figure 5 shows the calculated formation energies of the oxygen vacancy and Ti interstitial of rutile $TiO_2$ for both O-poor and O-rich conditions in sX. For $V_O^0$ in the O-rich condition, the calculated formation energy is +5.7 eV, correspnding to +1.9 eV for the O-poor condition. The transition energy levels correspond to the Fermi energy where the charge $q$ and $q'$ defect states have the same formation energies. The calculated transition energy of $V_O^0$ / $V_O^{2+}$ in sX is ~ 0.7 eV below the conduction band minimum $E_C$. Thus, the O vacancy is a deep defect. We see that the different charge states of the O vacancy all cross at 0.7 below Ec, so that the effective corrleation energy U of this defect is about 0. Similarly, U for the Ti interstitial is close to 0 in our sX results. Nevertheless, this does not stop the paramagnetic states of these defects being observed.

Fig. 6 shows the partial density of states (DOS) of the defect sites, for the three charge states $V_O^0$, $V_O^+$ and $V_O^{2+}$. In sX, the oxygen vacancy gives rise to a defect state lying well inside the band gap, as been found in the DOS for $V_O^0$ and $V_O^+$. The energy of the gap state is around 0.7 eV below the conduction band edge, agrees well with the experiments [8]. For the $V_O^0$, the defect state is occupied by two electrons; for the $V_O^+$, it is occupied by one electron. Hence there is no gap state for $V_O^{2+}$ because the excess electrons at the vacancy site have been ionised.

Figure 7 shows the charge density map of the defect state for $V_O^0$ and $V_O^+$. The neutral vacancy is S=1 with the electrons localised on two of the three adjacent Ti atoms next to the vacancy along the $O_z$ axis, rather than on the vacancy itself. In the positive vacancy the unpaired electron is also in a gap state mainly localized on the two adjacent Ti atoms along the $O_z$ axis. Both these are consistent with the EPR data [14].

We found that the localisation can also depend on the Ti pseudopotential. We also used Ti pseudopotential generated by the OPIUM method [53] to study the oxygen vacancy defect in rutile $TiO_2$. Except for the pseudopotential, all other settings such as $k_{TF}$ were the same. The resulting DOS also has a defect state in the gap, but the state for $V_O^0$ is localised on the $d_{xy}$ orbital, which is contrary to experimental findings. This shows the subtle effects that can occur in this borderline deep state.

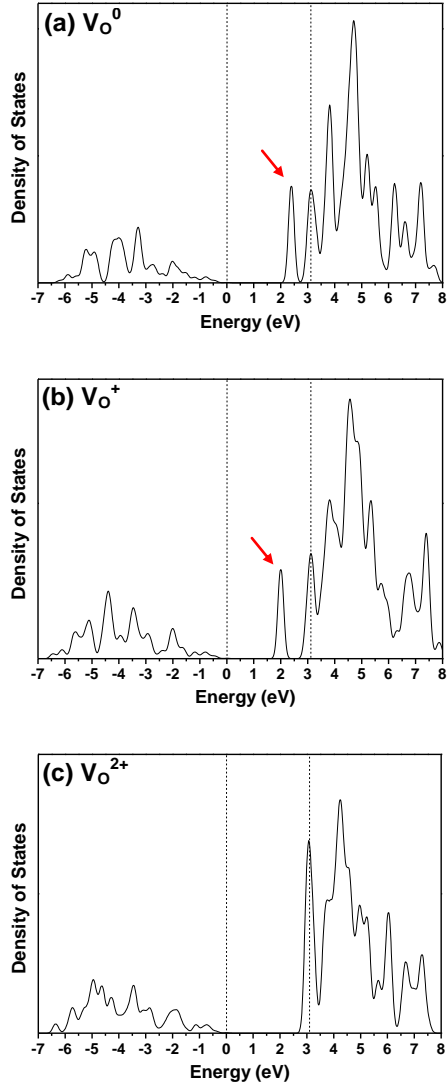

FIG. 6. The partial density of states (PDOS) of the Ti atom nearby the vacancy site for rutile $TiO_2$ in the sX method, (a) $V_O^0$ (b) $V_O^+$ (c) $V_O^{2+}$. The arrows indicate the defect states within the band gaps. The vertical dash lines denote the edge of valence band and conduction band. The top of the valence band is set to be the zero energy.

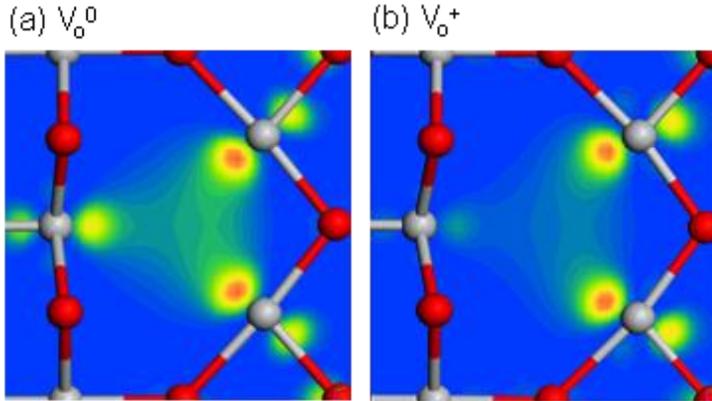

FIG. 7. (Color online) Charge density contours for the oxygen vacancy site of rutile $TiO_2$, (a) $V_O^0$ (b) $V_O^+$, calculated by the sX functional.

### D. Titanium interstitial

Ti interstitial ($Ti_i$) is the other key defect in reduced titania. It occupies an octahedral site surrounded by six O atoms as shown in Figure 2(b). The Ti interstitial has charge states $Ti_i^0$ to $Ti_i^{4+}$. After atomic relaxation by the sX functional for the neutral charge state ($Ti_i^0$), we found the inserted Ti atom pushes away the adjacent O atoms, forming a longer bond length at the octrahedral site, as shown in Figure 8. The relaxed Ti-Ti distance is 2.59 Å while the original distance is 2.28 Å. There are two different Ti-O length in the lattice, one is from 1.65 Å expands to 1.99 Å, the other is from 2.23 Å slightly shorten to 2.02 Å.

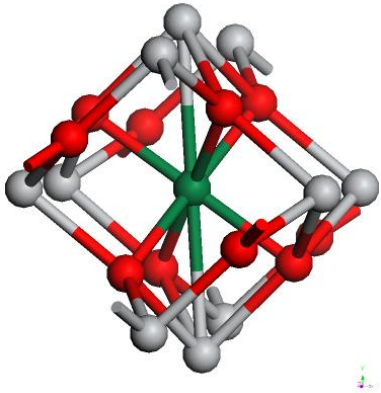
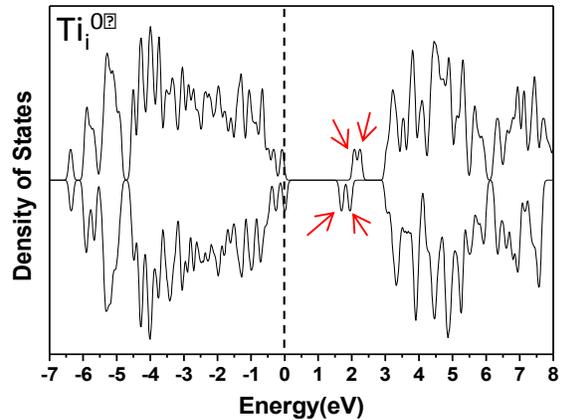

FIG. 8. (Color online) Equilibrium geometry of $Ti_i^0$ in rutile $TiO_2$ calculated by sX. Ti interstitial is shown green.

FIG. 9. The density of states (DOS) of $Ti_i^0$ in rutile $TiO_2$ calculated by the sX method. The red arrows indicate the gap states.

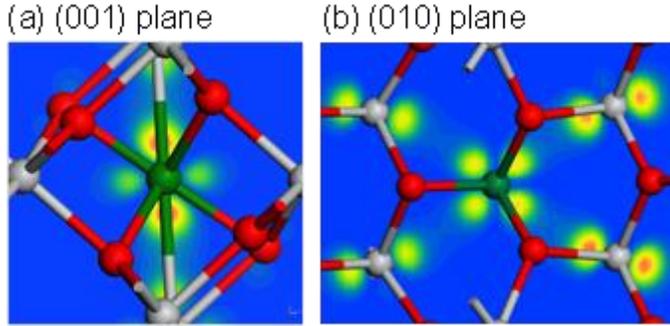

FIG. 10. (Color online) The charge density contours of the four in-gap defect states of the neutral Ti interstitial, calculated by the sX functional. (a) and (b) are (001) and (010) plane repectively. The interstitial Ti atom is shown as a green sphere.

Figure 9 shows the density of state of $Ti_i^0$ in rutile $TiO_2$. It is worth to mention that spin-polarisation is necessary to accurately describe the defect states of the Ti interstitial. We performed the non-spin polarised sX computation and found that only two peaks appeared near the conduction band minimum for $Ti_i^0$. We then applied the spin-polarisation, and the DOS now shows four defect states lying in the band gap at 0.7 ~ 1.3 eV below the conduction band edge, which is consistenmt with the experimental data [8-12]. Thus, sX also gives the $Ti_i$ as a deep state. From the charge density contours of the defect states in Figure 10, we found that the trapped electrons of the Ti interstitial were fully localised on the Ti 3$d$ orbitals at the inserted and adjacent Ti atoms. It supports the observation that the $Ti^{3+}$ ion is formed by Ti interstitial. Finazzi et al. [54] also noted the need for spin polarised calculation for Ti interstitial in their B3LYP calculations. Their non-spin-polarised B3LYP calculations on $Ti_i^0$ obtained only a doubly occupied state with higher total energy, that does not account for the EPR data.

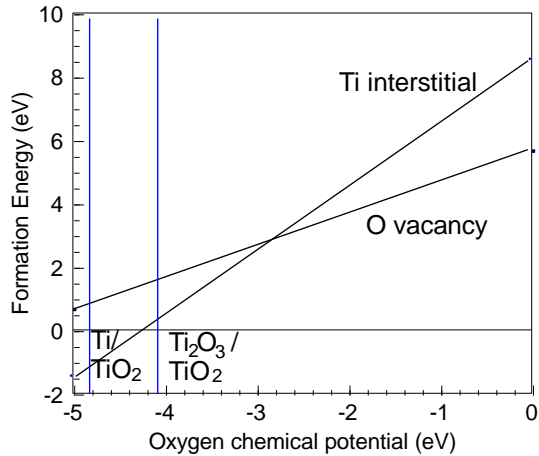

Fig. 11. Formation of energy of the neutral O vacancy and neutral Ti interstitial as a funciton of O chemcical potential, showing the cross-over of defect stabiltiy at chemical potentials below −2.8 eV.

Morgan and Watson [27,30] studied the oxygen vacancy formation energy using the GGA+U approach. In their result, the formation energy for the neutral charge state is 3.66 eV. Our formation energy of oxygen vacancy calculated by the sX functional is smaller than that calculated by GGA+U. In addition, Janotti et al. [37] used the HSE hybrid functional to obtain ~1.8 eV formation energy for the $V_O^0$. It is interesting that the formation energy value for $V_O^0$ is within ± 0.1 eV for both hybrid functionals, sX and HSE, but HSE predicts the oxygen vacancy as a shallow donor [37], whereas sX finds it to be a deep defect.

Fig 11 compares the formation energy of the neutral O vacancy and neutral Ti interstitial defects in sX as a function of the O chemical potential. This shows that the O vacancy is the more stable defect at higher $\mu_O$ but that the Ti interstitial becomes the more stable defect for $\mu_O$ below $-2.8$ eV. This does rationalise the various arguments in favor of the O vacancy being dominant on $TiO_2$ surfaces [9-13,18-19], whilst the Ti interstitial can contribute in the Ti-rich situation. Note that our cross-over $\mu_O$ value differs from that of Morgan and Watson [30] because their GGA+U method gave poorer values for the heat of formation of $TiO_2$. Interestingly, for the O-poor condition of $Ti_i^0$, the formation energy calculated by sX is 0.35 eV, less than for the O vacancy. Just below this, the $Ti_2O_3$ phase becomes the more stable.

These calculations will be extended to other donor systems such as interstitial hydrogen, which forms a localised but relatively shallow donor state in $TiO_2$ [35,55,56].

### E  Oxygen interstitial

The oxygen interstitial is the principle oxygen-excess defect. Its neutral configuration is the most stable across most of the range of Fermi energies (Fig 12). In this configuration, the interstitial O forms a dumbell O-O bonded ion with a lattice oxygen. In the -1 configuration, it also forms a dumbell, but with a longer O-O bond, while in its $-2$ structure the interstitial forms a separate ion. The O interstitial is able to show all its charge states in our sX calculation. Morgan and Watson [28] also studied the O interstitial, but because the GGA+U method is not able to open up the $TiO_2$ gap correctly, being a clsoed shell system, they could not access the negative charge states so easily. The O interstitial here behaves similarly to that in $HfO_2$ [57].

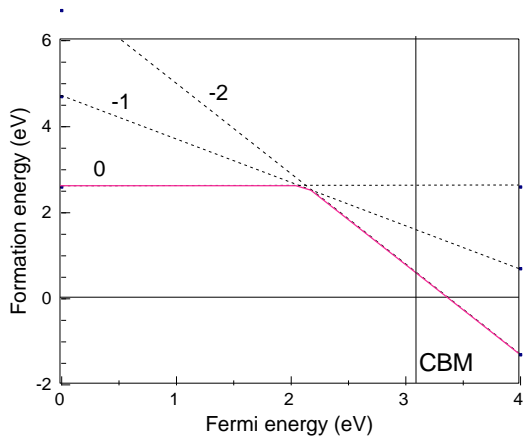

Fig. 12. Formation energy for the Oxygen interstitial in O-rich conditions.

### IV. CONCLUSIONS

We have calculated the oxygen vacancy and titanium interstitial in rutile $TiO_2$ using the sX hybrid functional. The sX method gives a band gap of 3.1 eV, close to the experimental value. The oxygen vacancy gives rise to localised defect states in the band gap, 0.7 eV below the conduction band, and their charge density are strongly localised on two of the three adjacent Ti atoms. The Ti interstitial also gives localised gap states, at 0.5-1.0 eV below the conduction band edge with their charge density localised on the interstitial and adjacent Ti atoms. The formation energies of the neutral oxygen vacancy and neutral Ti interstitial in O-poor condition are similar, at 1.9 eV and 0.35 eV respectively. In O-rich limit, the oxygen vacancy is easier to form compared to the Ti interstitial.

References
*jr214@cam.ac.uk
1. U. Diebold, Surf. Sci. Rep. **48**, 53 (2003).
2. T. L. Thompson and J. R. Yates, Chem. Rev. **106**, 4428 (2006).
3. A. Fujishima, X. Zhang, D. A. Tryk, Surf. Sci. Rep. **63**, 515 (2008)
4. M. Gratzel, Nature **414**, 338 (2001)
5. J. Robertson, Rep. Prog. Phys. **69**, 327 (2006).
6. R. Waser, R. Dittmann, G. Staikov, K. Szot, Adv. Mats. **21**, 2632 (2009).
7. S. G. Park, B. Magyari-Kope, and Y. Nishi, Phys. Rev. B **82**, 115109 (2010).
8. V. E. Henrich, G. Dresselhaus, and H. J. Zeiger, Phys. Rev. Lett. **36**, 1335 (1976).
9. R. L. Kurtz, R. Stockbauer, T. E. Madey, E. Roman, and J. L. Desegovia, Surf. Sci. **218**, 178 (1989).
10. M. Nolan, S. D. Elliott, J. S. Mulley, R. A. Bennett, M. Basham, and P. Mulheran, Phys. Rev. B **77**, 235424 (2008).
11. W. Gopel, J. A. Anderson, D. Frankel, M. Jaehnig, K. Phillips, J. A. Schafer, and G. Rocker, Surf. Sci. **139**, 333 (1984).
12. M. A. Henderson, W. S. Epling, C. H. F. Peden, and C. L. Perkins, J. Phys. Chem. B **107**, 534 (2003).
13. Z. M. Zhang, S. P. Jeng, and V. E. Henrich, Phys. Rev. B **43**, 12004 (1991).
14. S. Yang, L. E. Halliburton, A. Manivannan, P. H. Bunton, D. B. Baker, M. Klemm, S. Horn, and A. Fujishima, Appl. Phys. Lett. **94**, 162114 (2009).
15. S. Wendt, et al., Science **320**, 1755 (2008).
16. S. A. Chambers, S. H. Cheung, V. Shutthanandan, S. Thevuthasan, M. K. Bowman, A. G. Joly, Chem. Phys. Lett. **339**, 27 (2007)
17. K. Mitsuhara, H. Okumura, A. Visikovskiy, M. Takizawa, and Y. Kido, J. Chem. Phys. **136**, 124707 (2012).
18. C. M. Yim, C. L. Pang, and G. Thornton, Phys. Rev. Lett. **104**, 036806 (2010).
19. P. Kruger, J. Jupille, S. Bourgeois, B. Domenichini, A. Verdini, L. Floreano, and A. Morgante, Phys. Rev. Lett. **108**, 126803 (2012).
20. M. Ramamoorthy, R. D. King-smith, and D. Vanderbilt, Phys. Rev. B **49**, 7709 (1994).
21. E. Cho, S. Han, H. S. Ahn, K. R. Lee, S. K. Kim, and C. S. Hwang, Phys. Rev. B **73**, 193202 (2006)
22. J. M. Sullivan, S. C. Erwin, Phys. Rev. B **67**, 144415 (2003)
23. S. Na-Phattalung, M. F. Smith, K. Kim, M. H. Du, S. H. Wei, S. B. Zhang, S. Limpijumnong, Phys. Rev. B **73**, 125205 (2006)
24. P. Mori-Sanchez, A. J. Cohen, W. T. Yang, Phys. Rev. Lett. **100**, 146401 (2008)
25. G. Pacchioini, J. Chem. Phys. **128**, 182505 (2008)
26. S. L. Dudarev, G. A. Botton, S. Y. Savrasov, C. J. Humphreys, and A. P. Sutton, Phys. Rev. B **57**, 1505 (1998).
27. B. J. Morgan and G. W. Watson, Surf. Sci. **601**, 5034 (2007).
28. B. J. Morgan and G W Watson, Phys. Rev. B **80**, 233102 (2009)
29. B. J. Morgan and G. W. Watson, J. Phys. Chem. C **113**, 7322 (2009)
30. B. J. Morgan and G. W. Watson, J. Phys. Chem. C **114**, 2321 (2009)
31. J. Osorio-Guillen, S. Lany, A. Zunger, Phys. Rev. Lett. **100**, 036601 (2008)
32. D. C. Cronemeyer, Phys. Rev. **87**, 876 (1952); A. Amtout, R. Leonelli, Phys. Rev. B **51**, 6842 (1995)
33. S. Lany and A. Zunger, Phys. Rev. B **80**, 085202 (2009)
34. C. Di Valentin, G. Pacchioni, and A. Selloni, Phys. Rev. Lett. **97**, 166803 (2006).
35. C. Di Valentin, G. Pacchioni, and A. Selloni, J. Phys. Chem. C **113**, 20543 (2009).
36. J. Muscat, A. Wader, N. M. Harrison, Chem. Phys. Lett. **342**, 397 (2001).

37. A. Janotti, J. B. Varley, P. Rinke, N. Umezawa, G. Kresse, and C. G. Van de Walle, Phys. Rev. B **81**, 085212 (2010)
38. P. Deak, B. Aradi, T. Frauenheim, Phys. Rev. B **83**, 155207 (2011)
39. J. Heyd, G. E. Scuseria, and M. Ernzerhof, J. Chem. Phys. **118**, 8207 (2003).
40. J. Heyd, G. E. Scuseria, and M. Ernzerhof, J. Chem. Phys. **124**, 219906 (2006).
41. D. M. Bylander and L. Kleinman, Phys. Rev. B **41**, 7868 (1990).
42. A. Seidl, A. Gorling, P. Vogl, J. A. Majewski, and M. Levy, Phys. Rev. B **53**, 3764 (1996).
43. S. J. Clark and J. Robertson, Phys. Rev. B **82**, 085208 (2010).
44. K. Xiong, J. Robertson, M. C. Gibson, and S. J. Clark, Appl. Phys. Lett. **87**, 183505 (2005).
45. R. Gillen and J. Robertson, Phys. Rev. B **85**, 014117 (2012)
46. S. J. Clark, J. Robertson, S. Lany, and A. Zunger, Phys. Rev. B **81**, 115311 (2010).
47. R. Gillen and J. Robertson, unpublished
48. Y. Guo, S. J. Clark and J. Robertson, submitted to J. Phys. Conden. Mat. (2012)
49. S. J. Clark, M. D. Segall, C. J. Pickard, P. J. Hasnip, M. J. Probert, K. Refson, and M. C. Payne, Z. Kristallogr. **220**, 567 (2005).
50. S. Lany and A. Zunger, Phys. Rev. B **78**, 235104 (2008).
51. L. D. Finkelstein, et al., Phys. Rev. B **60**, 2212 (1999).
52. O. Kubaschewski, C. B. Alcock, 'Metallurgical Thermochemistry' (Elsevier 1979)
53. OPIUM pseudopotential package, http://opium.sourceforge.net
54. E. Finazzi, C. Di Valentin, and G. Pacchioni, J. Phys. Chem. C **113**, 3382 (2009).
55. P. W. Peacock, J. Robertson, Appl. Phys. Lett. **83**, 2025 (2003)
56. D. A. Panayotov, J. T. Yates, Chem. Phys. Lett. **436**, 204 (2007)
57. S. J. Clark, L. Lin and J. Robertson, Microelec. Eng. **88**, 1454 (2011)